\documentclass[prd,letterpaper,superscriptaddress,reprint]{revtex4}

\usepackage{amssymb}
\usepackage{amsthm}
\usepackage{amsmath}
\usepackage{amsbsy}
\usepackage{nicefrac}
\usepackage{wrapfig}
\usepackage{lineno}
\usepackage{float}
\usepackage[caption = false]{subfig}
\usepackage{natbib}
\usepackage{graphicx}
\include{rlFig}

\newcommand{\atUCLA}{Dept. of Physics and Astronomy, Univ. of California, Los Angeles, Los Angeles, CA 90095.}
\newcommand{\atOSU}{Dept. of Physics, Ohio State Univ., Columbus, OH 43210.}
\newcommand{\atUH}{Dept. of Physics and Astronomy, Univ. of Hawaii, Manoa, HI 96822.}
\newcommand{\atNTU}{Dept. of Physics, Grad. Inst. of Astrophys.,\& Leung Center for Cosmology and Particle Astrophysic\
s, National Taiwan University, Taipei, Taiwan.}

\newcommand{\atKU}{Dept. of Physics and Astronomy, Univ. of Kansas, Lawrence, KS 66045.}
\newcommand{\atWashU}{Dept. of Physics, Washington Univ. in St. Louis, MO 63130.}

\newcommand{\atUD}{Dept. of Physics, Univ. of Delaware, Newark, DE 19716.}
\newcommand{\atUCL}{Dept. of Physics and Astronomy, University College London, London, United Kingdom.}

\newcommand{\atJPL}{Jet Propulsion Laboratory, Pasadena, CA 91109.}
\newcommand{\atCCAP}{Center for Cosmology and Particle Astrophysics, Ohio State Univ., Columbus, OH 43210.}
\newcommand{\atChicago}{Dept. of Physics, Enrico Fermi Institute, Kavli Institute for Cosmological Physics, Univ. of C\
hicago , Chicago IL 60637.}

\newcommand{\atCalPoly}{Dept. of Physics, California Polytechnic State Univ., San Luis Obispo, CA 93407.}
\newcommand{\atIITT}{Dept. of Physics, Indian Institute of Technology, Kanpur, Uttar Pradesh 208016, India}
\newcommand{\atMEPHI}{National Research Nuclear University, Moscow Engineering Physics Institute, 31 Kashirskoye Highway, Rossia 115409}
\begin{document}

\title{Antarctic Surface Reflectivity Measurements from the ANITA-3 and HiCal-1 Experiments}

%
%

%
%

%
%


\author{P.~W.~Gorham}\affiliation{\atUH}


\author{P.~Allison}
\affiliation{\atOSU}
\affiliation{\atCCAP}

\author{O.~Banerjee}
\affiliation{\atOSU}

\author{J.~J.~Beatty}
\affiliation{\atOSU}
\affiliation{\atCCAP}

\author{K.~Belov}
\affiliation{\atJPL}

\author{D.~Z.~Besson}
\affiliation{\atKU}
\affiliation{\atMEPHI}

\author{W.~R.~Binns}
\affiliation{\atWashU}

\author{V.~Bugaev}
\affiliation{\atWashU}

\author{P.~Cao}
\affiliation{\atUD}

\author{C.~Chen}
\affiliation{\atNTU}

\author{P.~Chen}
\affiliation{\atNTU}

\author{J.~M.~Clem}
\affiliation{\atUD}

\author{A.~Connolly}
\affiliation{\atOSU}
\affiliation{\atCCAP}

\author{B.~Dailey}
\affiliation{\atOSU}

\author{P.~Dasgupta}
\affiliation{\atIITT}

\author{C.~Deaconu}
\affiliation{\atChicago}

\author{L.~Cremonesi}
\affiliation{\atUCL}

\author{P.~F.~Dowkontt}
\affiliation{\atUCLA}



\author{B.~D.~Fox}
\affiliation{\atUH}


\author{J.~Gordon}
\affiliation{\atOSU}



\author{B.~Hill}
\affiliation{\atUH}


\author{R.~Hupe}
\affiliation{\atOSU}

\author{M.~H.~Israel}
\affiliation{\atWashU}

\author{P.~Jain}
\affiliation{\atIITT}


\author{J.~Kowalski}
\affiliation{\atUH}

\author{J.~Lam}
\affiliation{\atUCLA}

\author{J.~G.~Learned}\affiliation{\atUH}

\author{K.~M.~Liewer}
\affiliation{\atJPL}

\author{T.C. Liu}
\affiliation{\atNTU}



\author{S.~Matsuno}
\affiliation{\atUH}


\author{C.~Miki}
\affiliation{\atUH}


\author{M.~Mottram}
\affiliation{\atUCL}

\author{K.~Mulrey}
\affiliation{\atUD}



\author{J.~Nam}\affiliation{\atNTU}

\author{R.~J.~Nichol}
\affiliation{\atUCL}

\author{A.~Novikov}
\affiliation{\atKU}
\affiliation{\atMEPHI}

\author{E.~Oberla}
\affiliation{\atChicago}


\author{S.~Prohira}
\affiliation{\atKU}

\author{B.~F.~Rauch}
\affiliation{\atWashU}




\author{A.~Romero-Wolf}\affiliation{\atJPL}

\author{B. Rotter}
\affiliation{\atUH}

\author{K.~Ratzlaff}
\affiliation{\atKU}

\author{J.~Russell}
\affiliation{\atUH}


\author{D.~Saltzberg}
\affiliation{\atUCLA}

\author{D.~Seckel}
\affiliation{\atUD}

\author{H.~Schoorlemmer}
\affiliation{\atUH}

\author{S.~Stafford}
\affiliation{\atOSU}

\author{J.~Stockham}
\affiliation{\atKU}

\author{M.~Stockham}
\affiliation{\atKU}

\author{B.~Strutt}
\affiliation{\atUCL}

\author{K.~Tatem}
\affiliation{\atUH}

\author{G.~S.~Varner}
\affiliation{\atUH}

\author{A.~G.~Vieregg}
\affiliation{\atChicago}

\author{S.~A.~Wissel}
\affiliation{\atCalPoly}

\author{F.~Wu}
\affiliation{\atUCLA}
\author{R.~Young}
\affiliation{\atKU}

\date{\today}

\begin{abstract}
The primary science goal of the NASA-sponsored ANITA project is measurement of
ultra-high energy neutrinos and cosmic rays, observed via radio-frequency
signals resulting from a neutrino- or cosmic ray- interaction with terrestrial matter
(atmospheric or ice molecules, e.g.). Accurate inference of the energies of
these cosmic rays requires understanding the transmission/reflection of radio wave
signals across the ice-air boundary.
Satellite-based measurements of Antarctic surface reflectivity, using a co-located
transmitter and receiver, have been performed more-or-less continuously for the last few 
decades. Our comparison of four
different reflectivity surveys, at frequencies ranging from 2--45 GHz
and at near-normal incidence,
yield generally consistent maps of high vs. low reflectivity, as a function of 
location, across Antarctica. 
Using the Sun as an RF source, and the ANITA-3 balloon borne radio-frequency 
antenna array as the
RF receiver, we have also measured the surface reflectivity over the interval 200-1000 MHz,
at elevation angles of 12-30 degrees. Consistent with our previous measurement using
ANITA-2, we find
good agreement, within systematic errors (dominated by antenna beam
width uncertainties) and across Antarctica,
with the expected reflectivity as prescribed by the Fresnel equations. To probe
low incidence angles, 
inaccessible to the Antarctic Solar technique and not probed by previous satellite
surveys,
a novel experimental approach
(``HiCal-1'') was devised.
Unlike previous measurements,
HiCal-ANITA constitute a bi-static
transmitter-receiver pair separated by hundreds of kilometers.
Data taken with HiCal, between 200--600 MHz shows 
a significant departure from the Fresnel equations, constant with frequency over that band, 
with the deficit
increasing with obliquity of incidence, which we attribute to the combined effects of possible surface roughness, surface grain effects, radar clutter and/or shadowing of the reflection zone due to Earth curvature effects.
We discuss the science implications of the HiCal
results, as well as improvements implemented for HiCal-2, launched in December, 2016.
\end{abstract}

\maketitle



\section{Introduction}
Within the last 30 years, the sub-field of ultra-high energy cosmic ray (UHECR; $E>10^{18}$ eV) astronomy has emerged as a vibrant experimental and theoretical sub-field within the larger field of particle astrophysics, comprising studies of both charged and neutral particles at macroscopic kinetic energies.
The physics interest in UHECR lies in understanding i) the nature of the cosmic accelerators capable of producing such enormously energetic particles at energies millions to billions of times higher than we are capable of producing in our terrestrial accelerators, ii) the details of the interaction of UHECR with the cosmic ray background, evident in the observed energy spectrum of cosmic rays as an upper `cut-off'\cite{GZK1,GZK2,GZK3}, or maximum observed energy, at approximately $10^{20}$ eV, and iii) correlations in the arrival directions of UHECR with exotic objects such as neutron stars, gamma-ray bursts (GRB), and active galactic nuclei (AGN). Experimentally, charged-particle UHECR astronomy
is currently dominated by two experiments -- the Southern Hemisphere Pierre Auger Observatory (PAO)\cite{AugerNIM2014} based in Malargue, Argentina and the Northern Hemisphere Telescope Array (TA)\cite{Abbasi:2014fya} based in Utah, USA. The construction of these observatories is very similar, based on a large number of ground detectors sampling the charged component of Extensive Air Showers (EAS) and with stations deployed over hundreds of square kilometers at $\sim$1 km spacing, coupled with a much smaller number of atmospheric nitrogen fluorescence detectors at sparser spacing having a much more restricted duty cycle, but individually capable of providing a more comprehensive image of atmospheric shower development. Within the last decade, the PAO has been complemented by an array of radio-wave antennas, capable of measuring the signal generated primarily by the separation of the charged particles comprising the down-moving air shower in the geomagnetic field\cite{AERAGroup2009,AERAEnergyPRL}. This technique has also demonstrated sensitivity to shower development and, therefore, the composition of the primary UHECR\cite{apel2014reconstruction}.

\section{Radiofrequency UHECR Detection with the ANITA Experiment}
Although originally purposed for detection of neutrinos, the ANITA-1 mission (2006) unexpectedly observed 14 extremely high-energy RF signals with a non-neutrino-like radiowave signal polarization (horizontal [HPol] vs. vertical [VPol], as expected for neutrinos) which traced back to the Antarctic surface beneath the balloon\cite{HooverNamGorham2010}. After considerable work, it was demonstrated that these events were the result of collisions of down-coming protons at ultra-high energies (corresponding to energies 10,000,000 times greater than the energy of particles accelerated in the Large Hadron Collider in Geneva, Switzerland) with atmospheric molecules. The RF signals produced in the collision, as the combined result of the so-called ``Askaryan effect'' resulting from the net charge excess acquired by the shower as it descends through the atmosphere, plus the ``geomagnetic'' signal resulting from separation of different charged species due to the Earth's magnetic field, were subsequently reflected off the Antarctic surface and back up to the ANITA balloon. Perhaps most striking was the discovery of two additional, steeply inclined events in the ANITA-1 sample entering the atmosphere nearly parallel to the Earth; these RF signals were observed directly rather than via surface reflection and had a signal polarity exactly opposite those observed via surface reflection, as expected. Within the last year, expanded analysis of the ANITA-1 event sample has uncovered one upcoming
event with signal polarity inconsistent with a reflection\cite{Gorham:2016zah}. Such an event does not readily
find a conventional explanation, and has been suggested as a possible $\nu_\tau$ candidate.

Recent analysis of the energy scale of those UHECR events, based on detailed modeling of the
signal generation\cite{Schoorlemmer:2015afa}, as well as fraction of primary cosmic ray-induced extensive air showers
reflected off the surface and back up the ANITA experiment indicate that the
efficiency-per-livetime for UHECR registration approaches that of the PAO and TA observatories.
 Accurate flux measurements require 
accurate inference of UHECR energies from their detected radio-wave signals. For ANITA's detection of protons, this means understanding surface effects on the reflected radio signal generated in an air-proton collision. Similarly, measurement of the energies of neutrinos colliding in-ice requires understanding of the surface-transmitted signal observed in-air by ANITA. 



\section{Experimental Measurement of Antarctic Surface Roughness and Reflectivity}
Several (interdependent) physical phenomena are likely responsible for the limited reflectivity of
the Antarctic surface. Wind-blown surface inhomogeneities are likely the primary determinant; wind
may also determine the typical locale-specific surface grain size. Sub-surface scattering effects
will also play some role, depending on frequency; to the extent that annual thin `crusts' of icy surface layers form as the temperature warms and then cools, volume and also layer scattering will both
contribute. Disentangling and quantifying all these features is an ongoing geophysical exercise.

\subsection{Satellite-based High Frequency Measurements}
\begin{figure}
\subfloat[Ka-band (26.5--40 GHz) Antarctic surface reflectivity, drawn from Envisat satellite data.]{\includegraphics[width=0.48\textwidth]{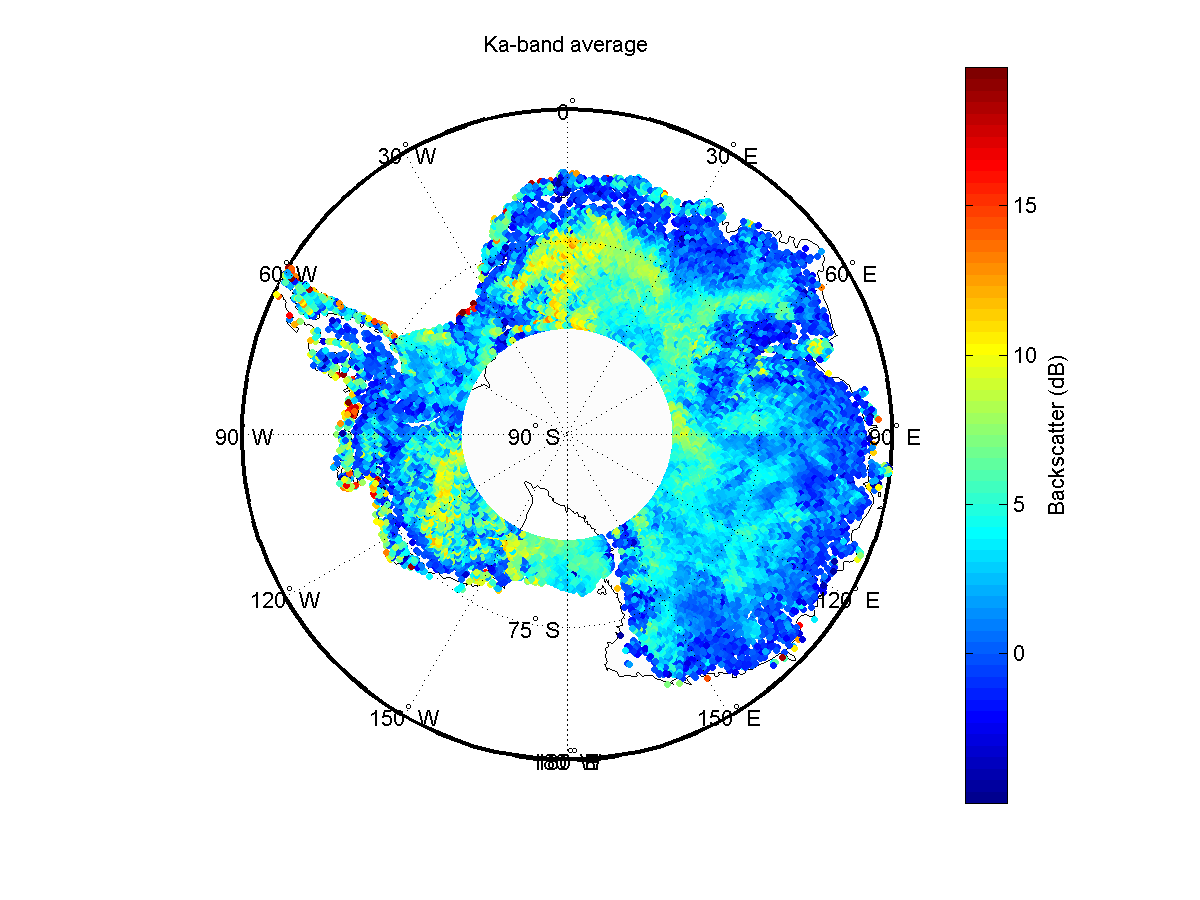}} 
\subfloat[Ku-band (12--18 GHz) Antarctic surface reflectivity, drawn from Envisat satellite data.]{\includegraphics[width=0.48\textwidth]{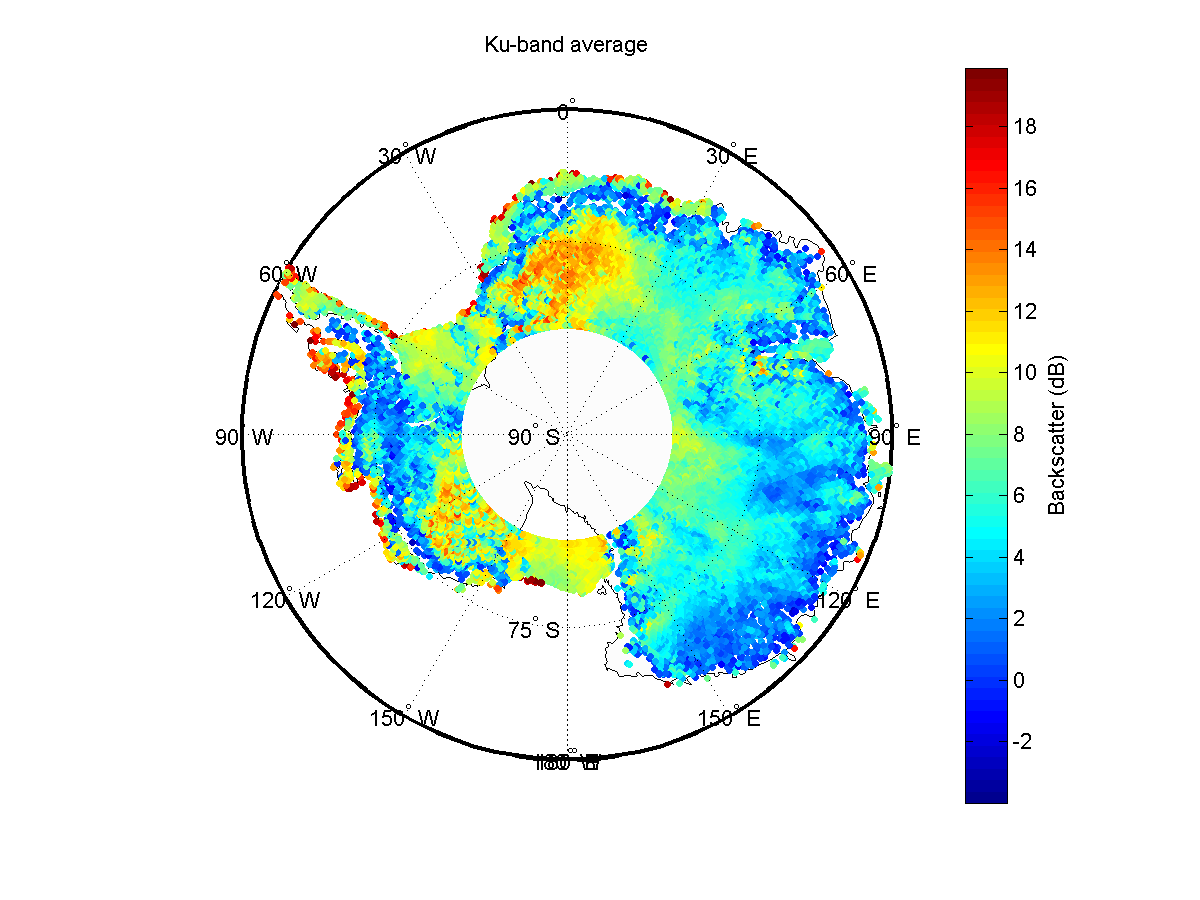}}    \\
\subfloat[S-band (2--4 GHz) Antarctic surface reflectivity, drawn from Envisat satellite data.]{\includegraphics[width=0.48\textwidth]{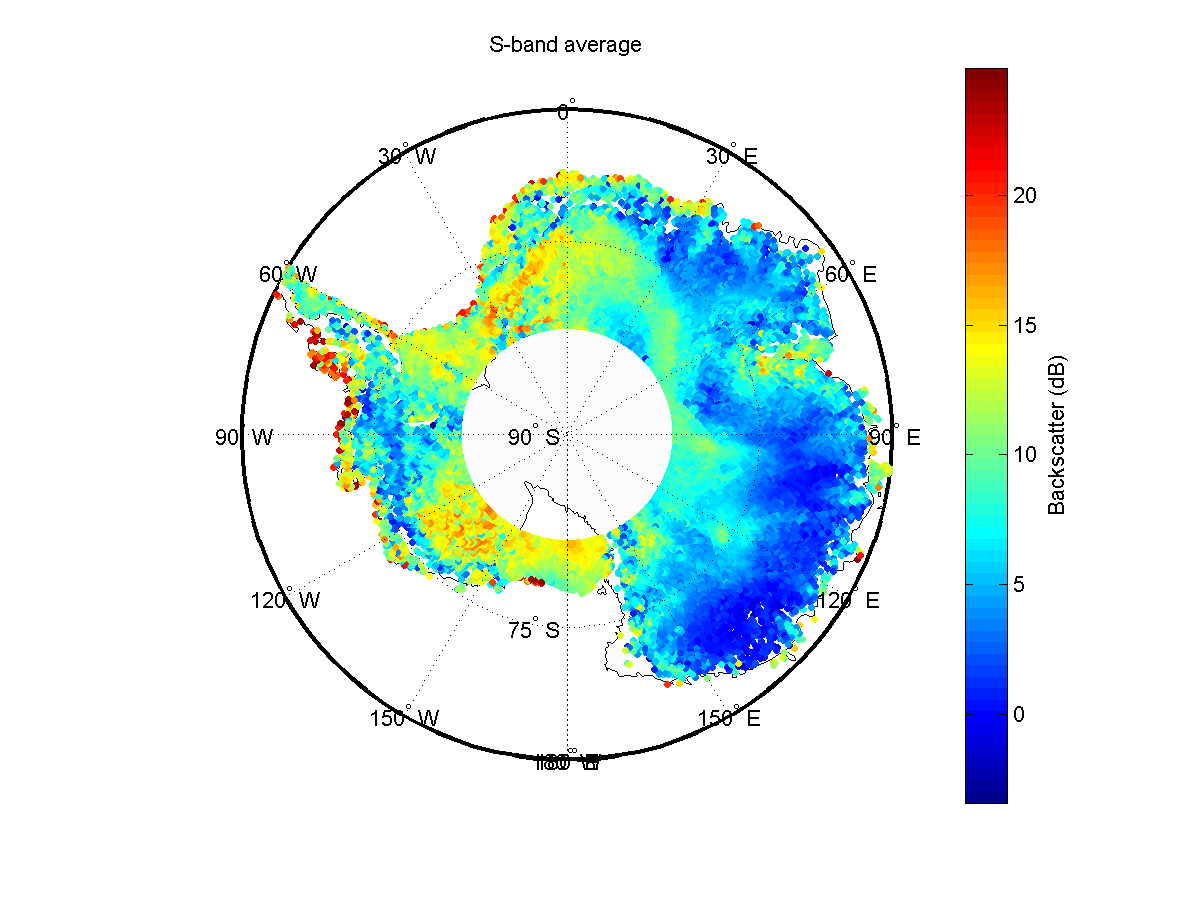}}
\subfloat[Correlation matrix between S-band, Ku-band, Ka-band reflecivities, as well as correlation with wind velocities.]{\includegraphics[width=0.48\textwidth]{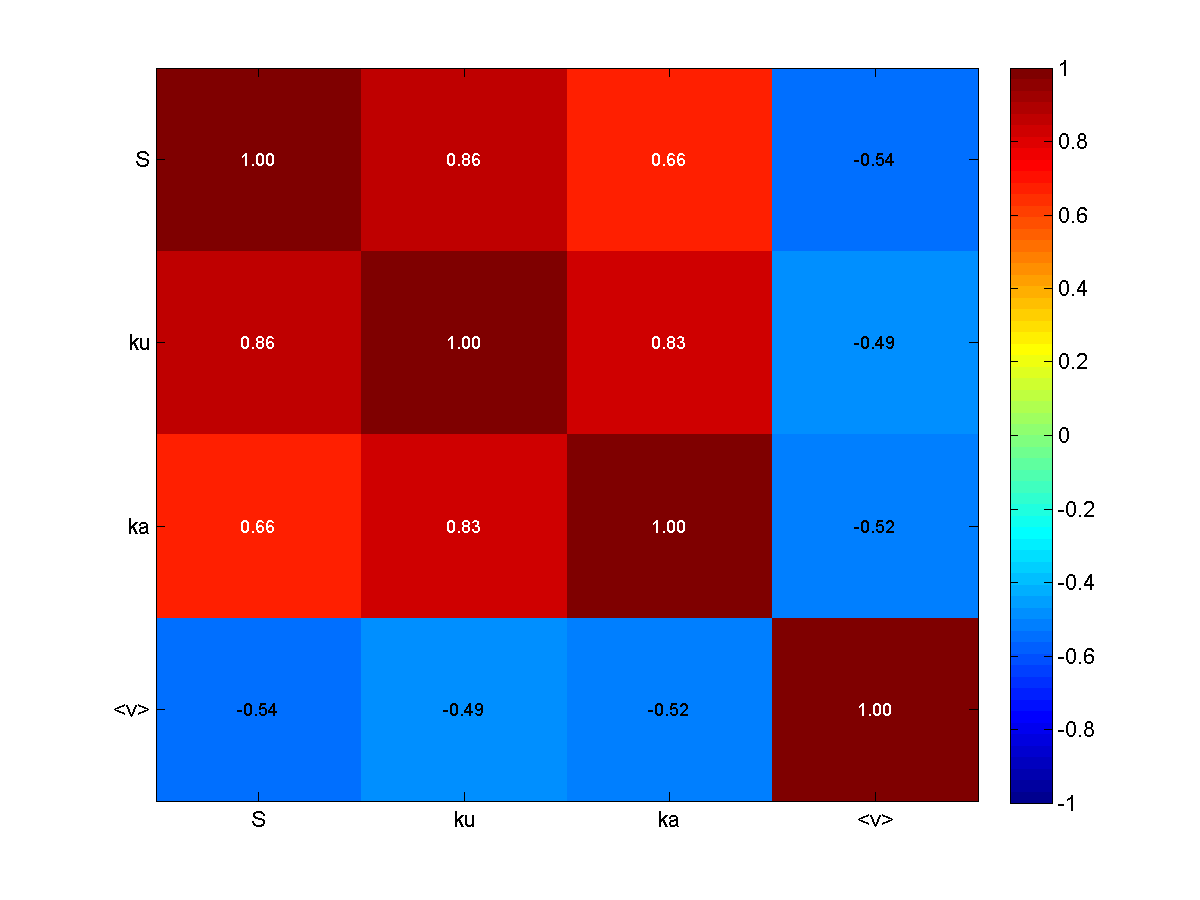}}
\caption{\it Compilation of satellite reflectivity data, over the frequency range 2--40 GHz, and correlation matrix.}
\label{fig:sats}
\end{figure}

At frequencies beyond 1 GHz, Antarctic surface reflectivity data have been taken
by the Envisat\cite{dubock2001envisat} and Aquarius\cite{neeck2005nasa} satellites. 
Figure \ref{fig:sats}
shows compilations of these data across the continent. Visually, the four bands show considerable similarity; interpretation of the L-band satellite data is somewhat complicated by the fact that those data have been taken in a variety of polarizations. We note that the smallest wavelength probed in these satellite surveys is $<$1 cm, suggesting that the surface is uniformly {\it incoherent} at wavelengths up to 30 cm.

We have considered the internal consistency of the higher-frequency satellite surveys with each other,
as well as the correlation between reflectivity and windspeed.
Figure \ref{fig:sats}d) summarizes the correlation between the three higher bands and also relative to wind velocity. We observe positive correlations in reflectivity across the continent for those three higher bands, indicating consistency in albedo measurements, as a function of position. We additionally, as expected, observe an anti-correlation between wind velocity and reflectivity, consistent with the expectation that higher wind velocities results in higher roughness and reduced albedo.



We conclude that, above 2 GHz, the satellite-based surveys are generally consistent and also
consistent with wind-driven surface effects reducing overall surface reflectivity. However, all these
measurements probe only surface scattering at normal incidence.


\subsection{Solar Measurements with ANITA-2 and ANITA-3}
\begin{figure}[htpb]
\includegraphics[width=0.64\textwidth]{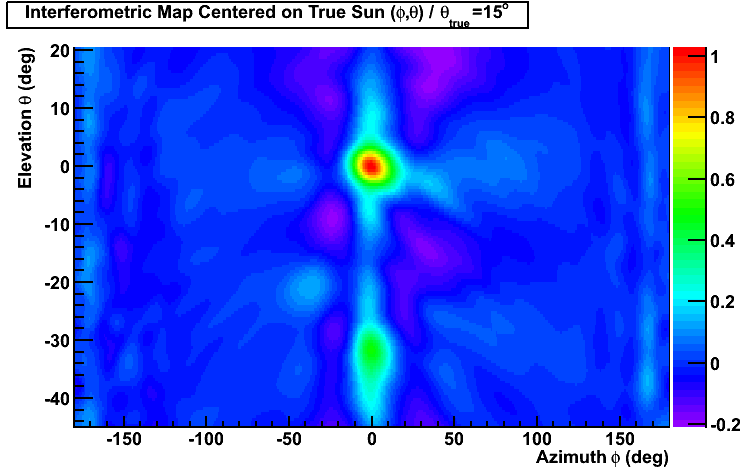}
\caption{\it Sample ANITA-3 sun-centered interferogram showing solar radio frequency image (at ($\phi,\theta\sim$0,0)) and reflection (at ($\phi,\theta)\sim(0,-32)$).}\label{fig:hSunCtr3}\end{figure}

We can probe surface roughness with radio wave receivers by measuring the ratio of the intensity of the surface-reflected radio-frequency Solar image to the Solar image observed directly by the balloon-borne ANITA experiment. That measurement can then be compared to the ratio expected for reflection off of a smooth surface (``specular'' reflection). 
By taking the ratio of the surface-reflected Solar RF power to the direct Solar RF power measured with ANITA, as a function of incident elevation angle relative to the surface, $\theta_i$, we can thus estimate the surface power reflection coefficients ${\cal R}(\theta_i)$. 
For $\theta_i>15^\circ$, our previous analysis found general consistency with the values of ${\cal R}(\theta_i)$ expected from the Fresnel equations\cite{hecht1974optics}, which prescribe the amount of signal power reflected at the interface between two smooth dielectrics given their indices of refraction, for both vertical- vs. horizontal-polarizations (``VPol'' and ``HPol'', respectively). At more glancing incident angles ($\theta_i < 15^ \circ$), the ANITA-2 data suggested slightly reduced signal strength compared to the expectation from the Fresnel equations, perhaps indicating that surface roughness effects are becoming increasingly apparent at oblique incidence angles.
Figure \ref{fig:hSunCtr3} shows one sample ANITA-3 HPol interferogram used to compile the reflection coefficients, as a function of incidence angle. In general, the ANITA-3 data follow the trend obtained from the ANITA-2 data.

\subsection{The HiCal Experiment}
The HiCal (High-altitude Calibration) balloon-borne transmitter was proposed to emulate the radio signals produced by UHECR and to
derive the effects of surface reflectivity and roughness, using ANITA as the receiver. 
To the extent that the generated HiCal signal matches the waveform expected from UHECR, HiCal triggers registered by
ANITA also can be used as a signal `template' in offline cosmic-ray search analysis.
In January, 2015, this technique was successfully prototyped 
with the HiCal-1 flight in Antarctica, tracking ANITA-3.  
In this scheme, a `trailer' balloon (``HiCal-1''), comprising an in-air RF transmitter emitting high amplitude signals measured by ANITA both directly (``D''), as well as in their surface reflection (``R''), is launched in proximity to the ANITA flight path.
The ratio of the measured ANITA amplitude from a surface-reflected HiCal signal relative to a directly-received signal, 
over a wide range of incidence angles numerically defines the reflectivity; the short $\sim$10$\mu$s separation time of these two signals gives a unique, and easily recognizable signature in the ANITA data sample. Knowing the GPS coordinates of both ANITA as well as HiCal at any given time, we can calculate the expected time difference between the reflected and direct RF signals. At these large separations, inclusion of Earth curvature effects, as well as the $\sim$2-3 km elevation of the Antarctic plateau at the putative intervening RF reflection point, are critical. 


Note that the HiCal-ANITA transmitter-receiver pair represent a bi-static radar configuration, for which the dependence of the received signal on the radar beam direction relative to the sastrugi alignment can be opposite that expected for monostatic radar. In the case of sastrugi aligned transverse to the radar beam, monostatic devices such as satellites directly measure the enhanced radar backscatter, while in the bi-static configuration, signal can be affected by such things as, e.g., local shadowing.



\subsection{HiCal hardware}
\subsubsection{Payload Schematic and Transmitter}
\begin{wrapfigure}{R}{0.5\textwidth} 
\centerline{\includegraphics[width=0.48\textwidth,angle=0]{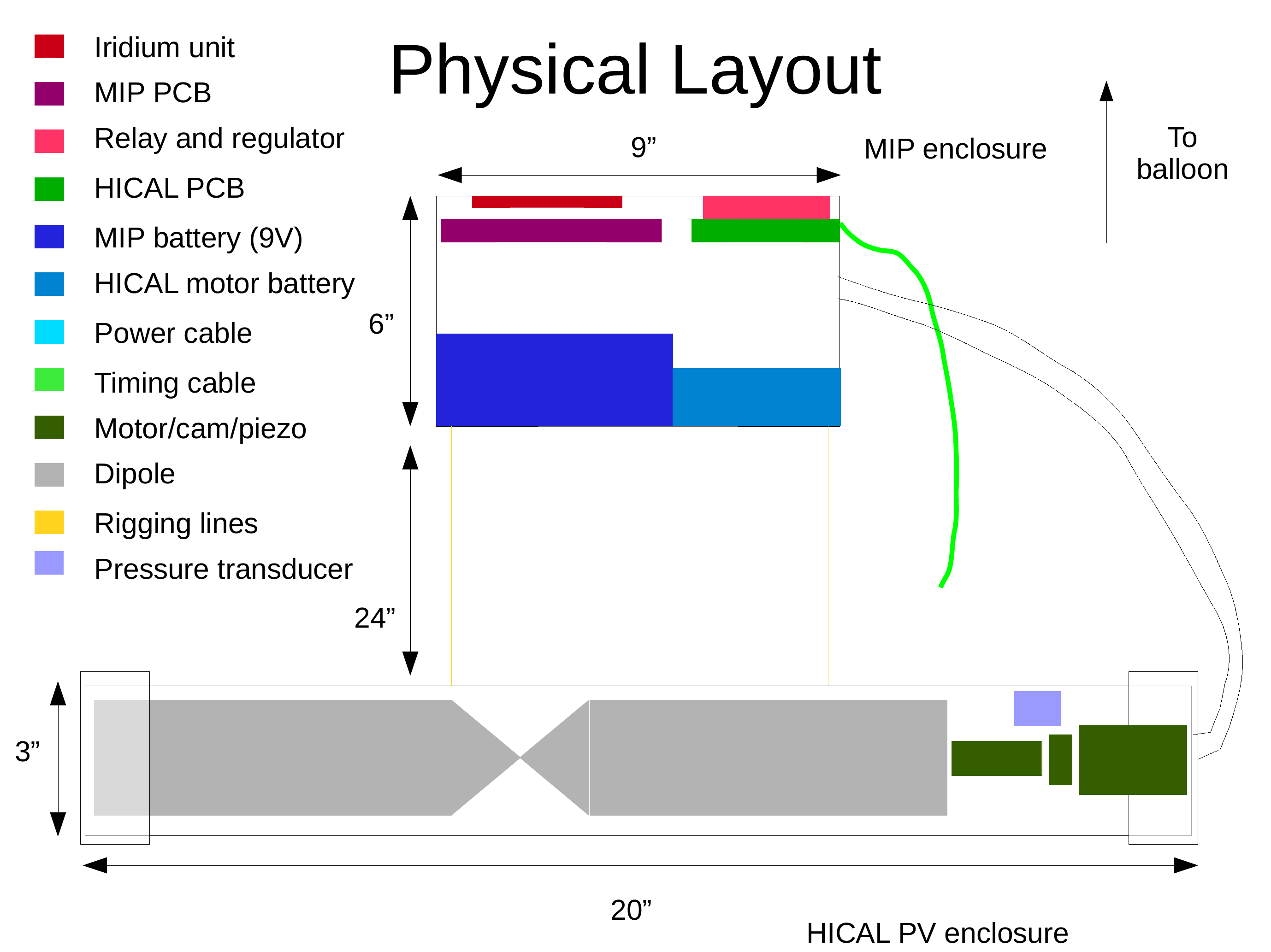}}
\caption{\it HiCal-1 payload schematic, showing electronics box, comprising micro-instrumentation package (MIP) compartment, including GPS and MIP battery, and HiCal compartment, plus RICE dipole transmitter below. MIP is supplied by Columbia Scientific Balloon Facility (CSBF), funded by NASA.}\label{fig:schematic}
\end{wrapfigure}

The HiCal-1 transmitter is based on a small ceramic piezo-electric. 
Such devices translate the mechanical energy of impact of a solid `actuator' with a piezo ceramic into a 
$\sim$10 nanosecond-duration burst of electrical energy, and are capable of generating kiloVolt-scale radio-frequency signals. The full HiCal payload consists of three sub-elements, schematically illustrated in Figure 
\ref{fig:schematic}. The ``Micro-Instrumentation Package'' (MIP) is a NASA standard for sub-orbital missions, and contains the hardware for communications with the payload and control operations during flight (telemetry), as well as GPS payload time and location information, which runs asynchronously relative to the HiCal triggers. Below the MIP, the ``actuator'' comprises a motor turning at a rep rate of approximately 0.33 Hz which drives a camshaft, designed to depress the spring-loaded piezo electric at the same 0.33 Hz frequency. Signals from the piezo are directed into the dipole antenna (built according to the RICE experiment's antenna specifications\cite{Kravchenko:2001id}), for which the feed point has been coated with anti-coronal paint to suppress possible arcing of the kiloVolt signals emitted by the piezo. 
Ultimately, given the enhanced arcing with which one must contend at the 38-km HiCal float altitude (5 mB pressure), a dedicated pressure vessel, constructed from lightweight ABS (Acrylonitrile-Butadiene-Styrene), was built to enclose the dipole and piezo in a sealed, 1000 mB environment. 
A second GPS board time stamps the RF signals being emitted by the dipole.

\subsubsection{Azimuthal Orientation Readout}
Since the emitted signal strongly depends on the orientation of the dipole transmitter relative to the direction of the ANITA balloon and since the payload can freely rotate during flight, an additional custom printed-circuit board also provides azimuthal orientation information (``HiCaz''). This board consists of 8 photodiodes spaced evenly around the circumference of the PC board, 
allowing determination of the azimuthal orientation relative to the instantaneous position of the sun in the sky, by interpolation of the measured photodiode voltages. After a slope correction, we obtain an azimuthal orientation resolution of approximately 1--2 degrees, which is smaller than the intrinsic uncertainty in the transmitter antenna directivity. 

\subsubsection{Signal Generation Details; Spark Gap Dependence}

\subsection{HiCal-1 flight details}
\begin{figure}[htpb]
\begin{minipage}{0.48\textwidth}
\includegraphics[width=1.\textwidth,angle=0]{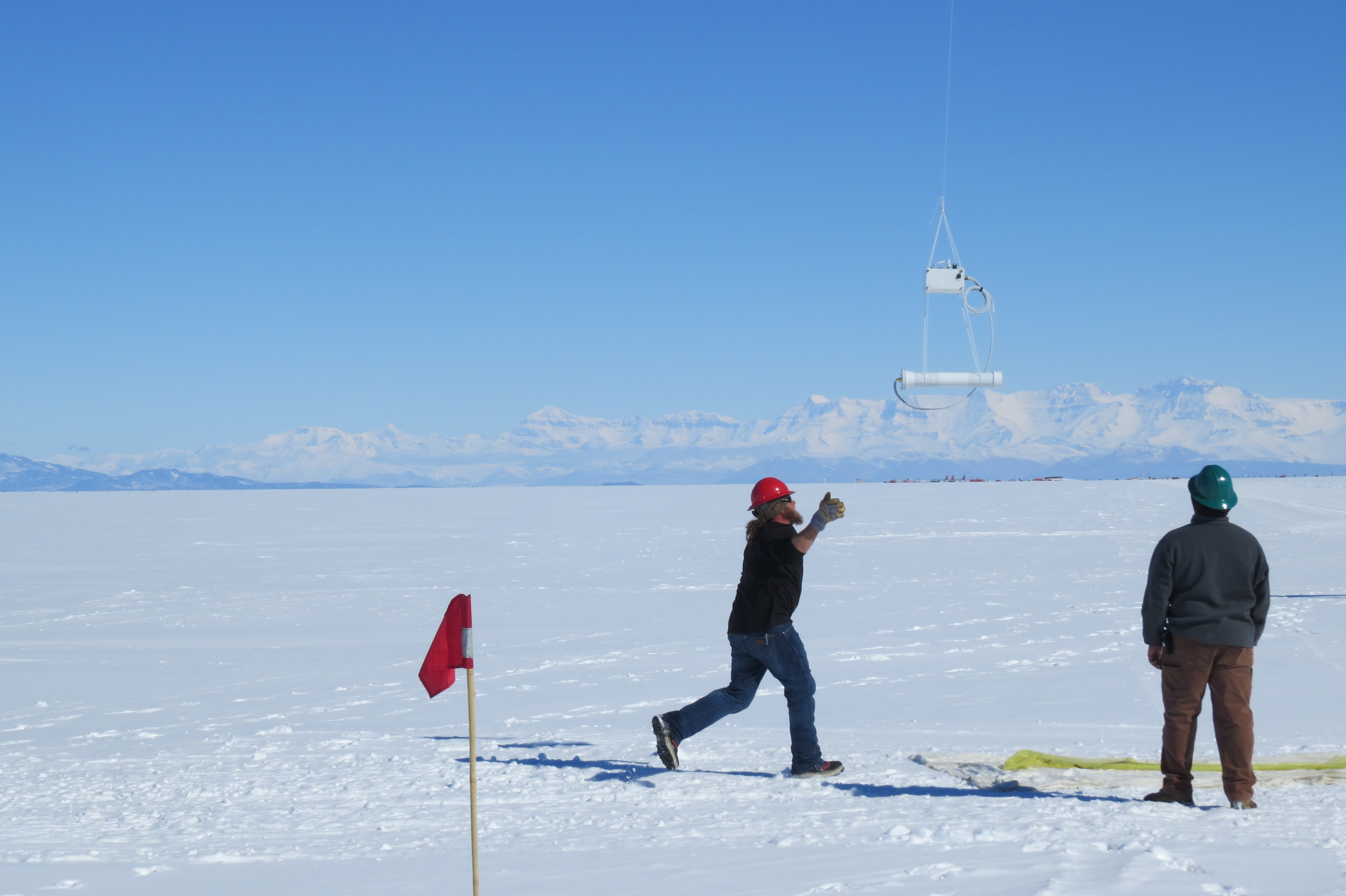}\caption{\it Zoom of HiCal payload directly following launch.}\label{fig:launch0}
\end{minipage}
\begin{minipage}{0.48\textwidth}
\includegraphics[width=1.\textwidth,angle=0]{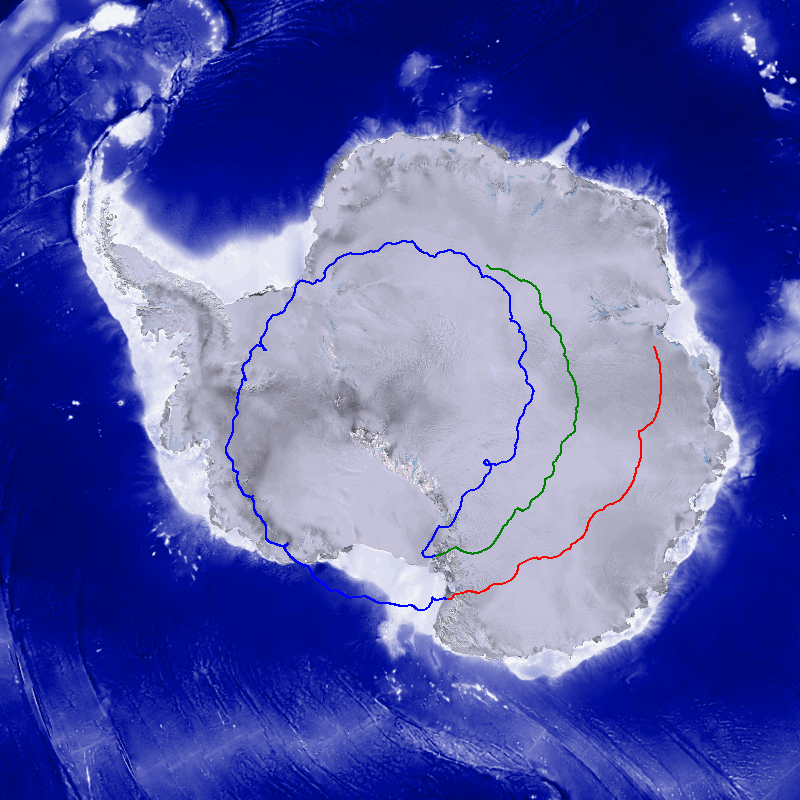}\caption{\it ANITA-3 flight track (blue for first orbit and red for second) overlaid with HiCal-1b (green).}
\label{fig:csbf}
\end{minipage}
\end{figure}

Four science projects were approved for launch for the 2014-15 NASA Long-Duration Balloon campaign in Antarctica.
These were the ANITA-3 project, the Super Pressure Balloon COSI\cite{chiu2015upcoming} project, the SPIDER\cite{crill2008spider} project, and HiCal. 
The first HiCal launch attempt (``HiCal-1a''), on December 19, 2014 was unsuccessful. Insufficient lift, which resulted in a balloon which failed to ascend, was compounded by erratic piezo output pulses. A second launch (Fig. \ref{fig:launch0}), using back-up HiCal hardware 
occurred 
on January 6, 2015 (``HiCal-1b'') during ANITA-3's return after one circuit around the continent. 
Almost immediately after turning on the HiCal transmitter during ascent, ANITA began registering D triggers, the $\sim$700 km separation distance notwithstanding. We note that no clear R signals were observed during that time, indicating very small reflection coefficients at the near-glancing angles typical of ascent. After HiCal reached its 38-km float altitude, during those times when the transmitter was activated by the HiCal motor, signals continued to be recorded for the subsequent 48 hours (after which time the ANITA flight was terminated), at ANITA-HiCal separations between 650 and 800 km. The flight tracks of ANITA-3 (blue for first orbit and red for second orbit) is overlaid with that of HiCal-1b (shown in green) in Figure \ref{fig:csbf}  (reproduced from the CSBF/NASA website at http://www.csbf.nasa.gov/antarctica/ice.htm). 


In the initially telemetered ANITA data sample, three `doublet' events were quickly identified for which the time separation between the first pulse (the direct HiCal trigger D) and the second (the surface-reflected signal R) was of order 7.2 microseconds. Figure \ref{fig:dt} shows that these events agree excellently with calculations of the expected direct-reflected signal time delays based on the known HiCal-ANITA separation (640 km) at the time these events were recorded. 

In total, HiCal-1b produced $\sim$600 triggers observed by ANITA, many at ANITA/HiCal separation distances of $\sim$750 km, or 200 km further than the ANITA ground pulsers can be seen, 
due to Earth curvature effects. 

\subsection{HiCal science results}
\subsubsection{Time Characeristics of R and D signals}
\begin{figure}[htpb]
\begin{minipage}{3.5in}
\centerline{\includegraphics[width=1.\textwidth]{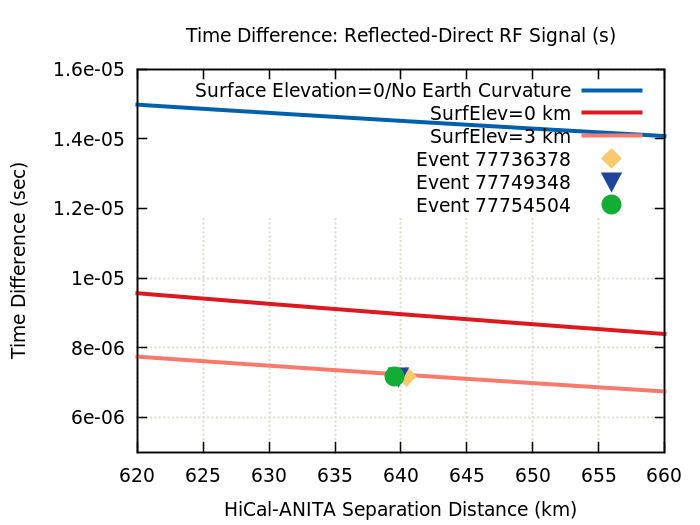}} \caption{\it Measured time difference between successive signals registered in telemetered sub-sample of ANITA-3 data (points) vs. expectation, given known GPS locations of transmitter and receiver (curves). Bottom curve, from HiCal MC simulations, corresponds to known 3 km Antarctic plateau elevation and includes Earth curvature effects.}\label{fig:dt} 
\end{minipage}
\begin{minipage}{3.5in}
\centerline{\includegraphics[width=1.\textwidth]{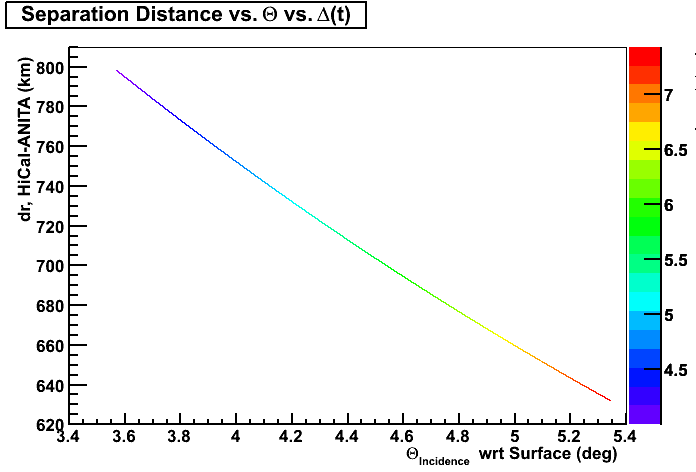}} 
\caption{\it Calculated incidence angle of Reflection signals relative to horizon (x-axis) and time difference between Direct and Reflected pulses (color scale, in microseconds) as a function of HiCal and ANITA separation distance (km; along y-axis).}\label{fig:HASepa} 
\end{minipage}
\end{figure}

The known separation distance between HiCal and ANITA, combined with elevation information of the 
Antarctic plateau, can be used to infer an expected time delay between registration of the reflected vs. the direct 
signals, as shown in Fig. \ref{fig:HASepa}. 
At typical separation distances of order 600-800 km, we expect time differences between R and D triggers
to be of order 7 microseconds. 

\begin{figure}[htpb]
\includegraphics[width=0.32\textwidth]{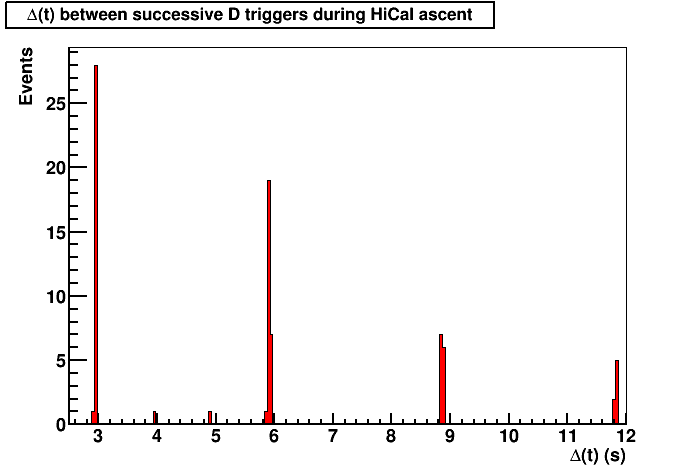}
\includegraphics[width=0.32\textwidth]{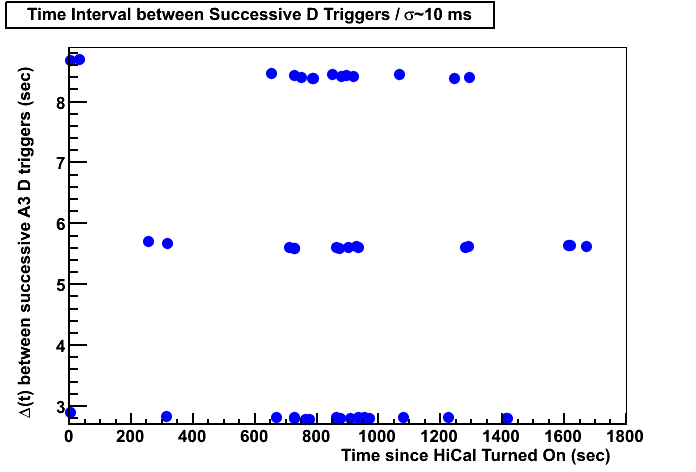}
\includegraphics[width=0.32\textwidth]{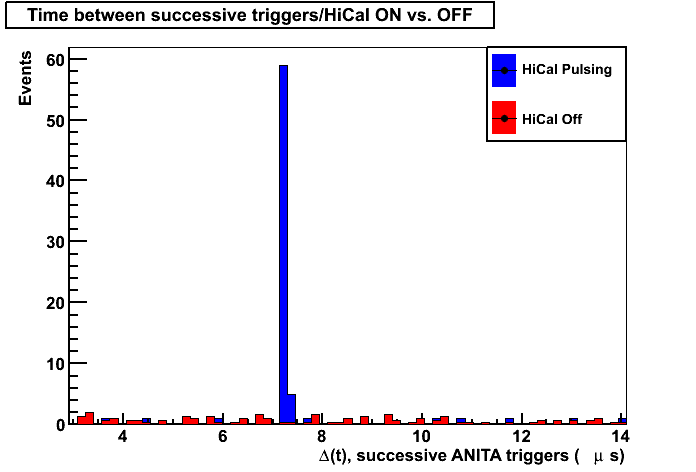} 
\caption{\it Left: Difference between successive D triggers, registered for ANITA runs 401 and 402, during HiCal-1b ascent. Center: Same, registered for ANITA run 413, for which the ANITA-HiCal separation distance was of order 680 km, and for which the HiCal motor was on (and therefore HiCal-1 presumably pulsing). We note the evident $\sim$3 second period of the transmitter pulsing; we also observe a large fraction of times when HiCal was nominally pulsing but no triggers registered at ANITA-3. We attribute at least some of these to cases where the HiCal transmitter antenna was in a geometrically disfavorable orientation relative to ANITA-3. Right: Time difference between successive registered HiCal-direct signals during one pulsing period.}\label{fig:HCmotorPeriod} 
\end{figure}

The camshaft actuator rotates with a period of approximately 3 seconds; Fig. \ref{fig:HCmotorPeriod} illustrates
the time between successive HiCal direct triggers registered offline by ANITA during ascent (left) and 
after reaching float (center). We note the clear presence
of a 3 second periodicity, although the fact that the time interval between successive triggers is often
6, 9, 12... seconds obviously indicates that several triggers are 'missing' from the ANITA-3 data stream.
This is due either to lower amplitude transmitter signals or, given the fact that the payload is rotating, an unfavorable
HiCal transmitter azimuthal orientation relative to the ANITA payload. HiCal-2 includes improved azimuthal information
to resolve this uncertainty. Figure \ref{fig:HCmotorPeriod} (right) shows the time difference between
successive ANITA-3 triggers, taken while HiCal-1b was pulsing, and showing the expected time difference
between R and D events.

\begin{figure}[htpb] 
\centerline{\includegraphics[width=0.6\textwidth]{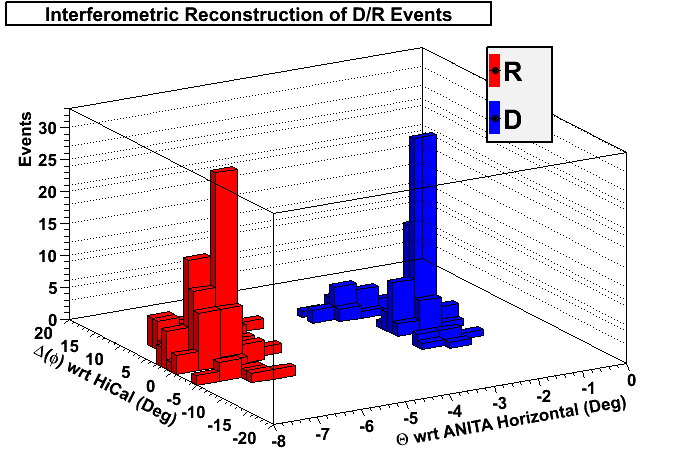}}
\caption{\it Angular reconstruction of HiCal direct and reflected events, in elevation ($\theta$) and azimuth ($\phi$). In these coordinates, $\theta$=0 corresponds to Horizontal relative to ANITA; $\theta\sim -6^o$ corresponds to the Earth horizon. Azimuth has subtracted out the known azimuthal location of the HiCal-1 transmitter at a given time.}\label{fig:DRrecon.png} \end{figure}

Figure \ref{fig:DRrecon.png} shows the angular reconstruction of the direct and reflected signals.
As expected, the direct signals emanate from a point slightly below horizontal, but above a tangent to the
Earth, whereas R signals appear to emanate from a point on the ice.

\subsubsection{Reflectivity Measurements}
\begin{figure}[htpb]
\begin{minipage}{3.25in}
\centerline{\includegraphics[width=1.\textwidth]{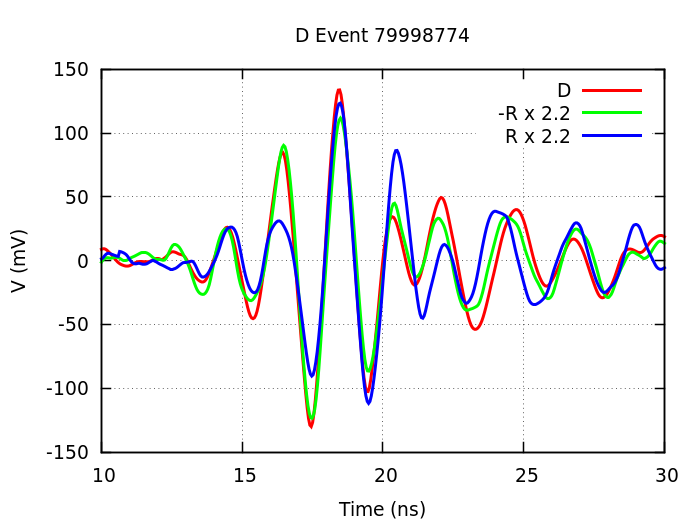}} 
\caption{\it Overlay of HiCal-1 Direct event 79998774 (red) overlaid with inverted event 79998775 (green) and non-inverted event 79998775 (blue), illustrating expected signal inversion for HiCal-1 reflected pulse.}\label{fig:79998774.png} \end{minipage}
\begin{minipage}{3.25in}
\centerline{\includegraphics[width=1.\textwidth]{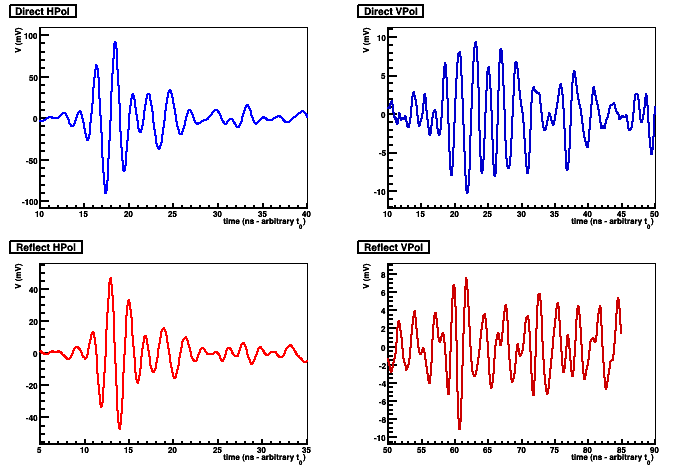}} 
\caption{\it Coherently summed, averaged waveforms for Direct HPol (upper left), Direct VPol (upper right), Reflected HPol (lower left) and Reflected VPol (lower right). Horizontal offset from zero, for all plots, is arbitrary. }\label{fig:DR_HV.png} \end{minipage}\end{figure}

The goal of HiCal is to obtain an estimate of the surface reflectivity for 
surface incidence angles less than
5 degrees. 
One of the signature features of signals reflecting off higher index-of-refraction
materials relative to direct signals
is the expected signal inversion. Figure \ref{fig:79998774.png} shows that
the inverted R signal gives, modulo an overall scale factor, a very good match to the directly-measured
signal. 
In fact, in the entire sample of doublet events consisting of an observed reflection as well as an observed direct signal, the cross-correlation for the reflected signal exceeds the cross-correlation of the direct event 100\% of the time.
The match of the signal shape also indicates that, to a good approximation, 
the frequency composition of the reflected
signal matches that of the direct signal.
Figure \ref{fig:DR_HV.png} shows the coherently summed, and averaged waveforms for direct vs. reflected
signals, again verifying the consistency of signal shape across the two samples.
We also note from Figure \ref{fig:DR_HV.png} that the direct VPol signal is reduced by approximately a factor of 10 in voltage (100 in power) relative to the HPol signal, indicating that the combined effects of cross-polarization broadcast from the transmitter dipole, plus cross-polarization response of the ANITA horn antennas, plus any possible vertically transmitted component resulting from the transmitter possibly being non-horizontal and having a non-zero polar angle offset, is reduced by approximately 20 dB relative to the expected HPol broadcast and received power.

In these measurements, and more critically for the Solar measurements, one must correct
for the different elevation angles of the direct vs. reflected signals. Since the ANITA antennas
are canted at a downwards angle of 10 degrees, reflected signals are closer to boresight than direct
signals. Consequently, the measured signal power for R vs. D must be multiplied by a factor
$exp(-(\theta_R+\theta_{Cant})^2/2\sigma_\theta^2(f))/
exp(-(\theta_D+\theta_{Cant})^2/2\sigma_\theta^2(f))$, with $\theta_R$ the elevation angle to
the reflection point, $\theta_D$ the elevation angle to the direct source, and
$\sigma_\theta$ the beam width (3 dB full-width-half-max/2.36) at a
given frequency, to obtain the actual reflected power ratio. 
\begin{figure}[htpb]
\includegraphics[width=1.\textwidth]{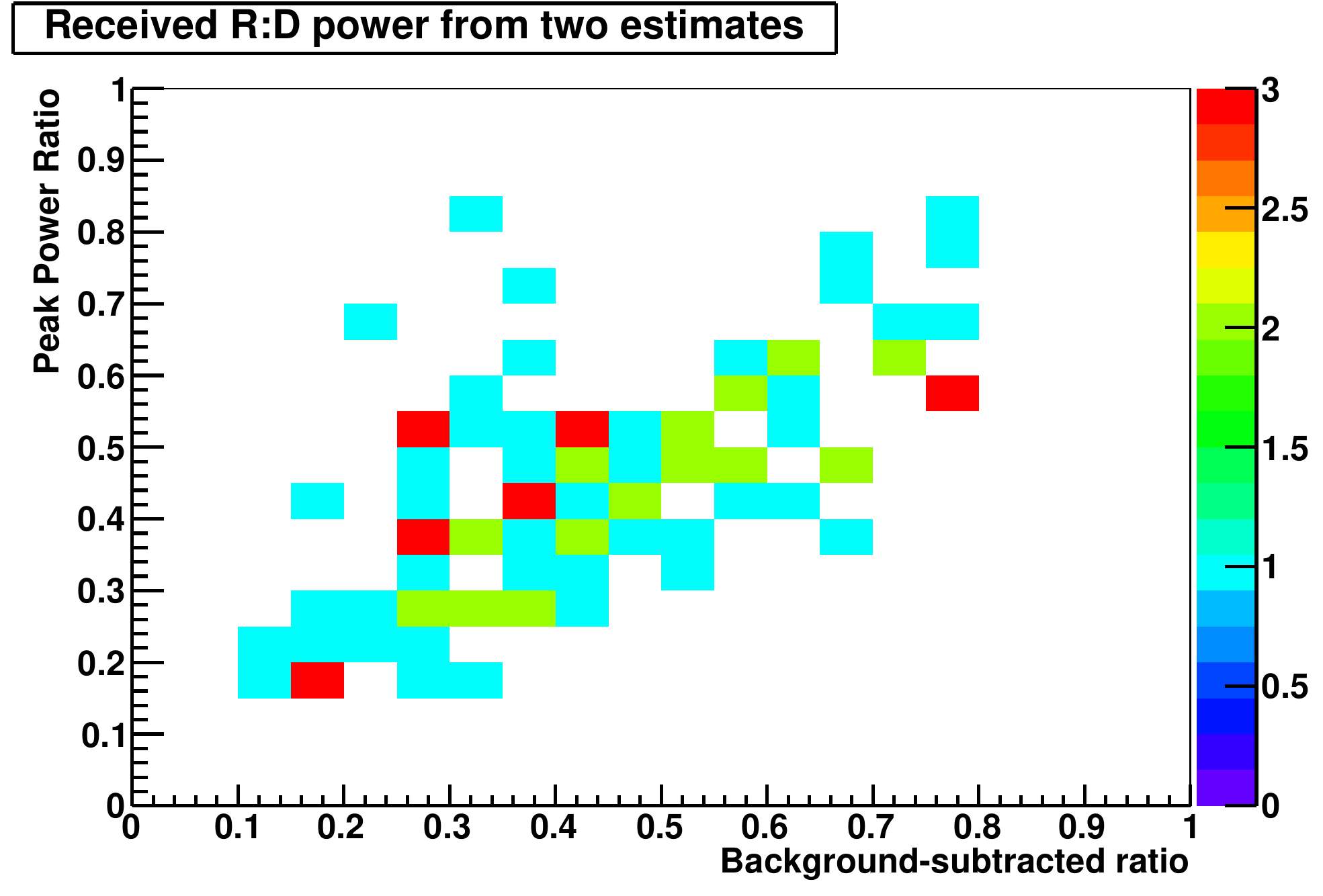}
\caption{\it Ratio of R:D signal strength using two different estimators.}\label{fig:DvsRscatterplot} 
\end{figure}
Figure \ref{fig:DvsRscatterplot} presents the raw ratio of R:D using two different estimators: in the
first, the R and D signal power estimated from the interferometric map are sideband subtracted, to
remove any DC offsets. In the second, the peak of the interferometric maps, for the pixels corresponding to the D and R signals are directly compared. The two estimates provide generally consistent measures of the reflectivity.

\begin{figure}[htpb]
\includegraphics[width=1.\textwidth]{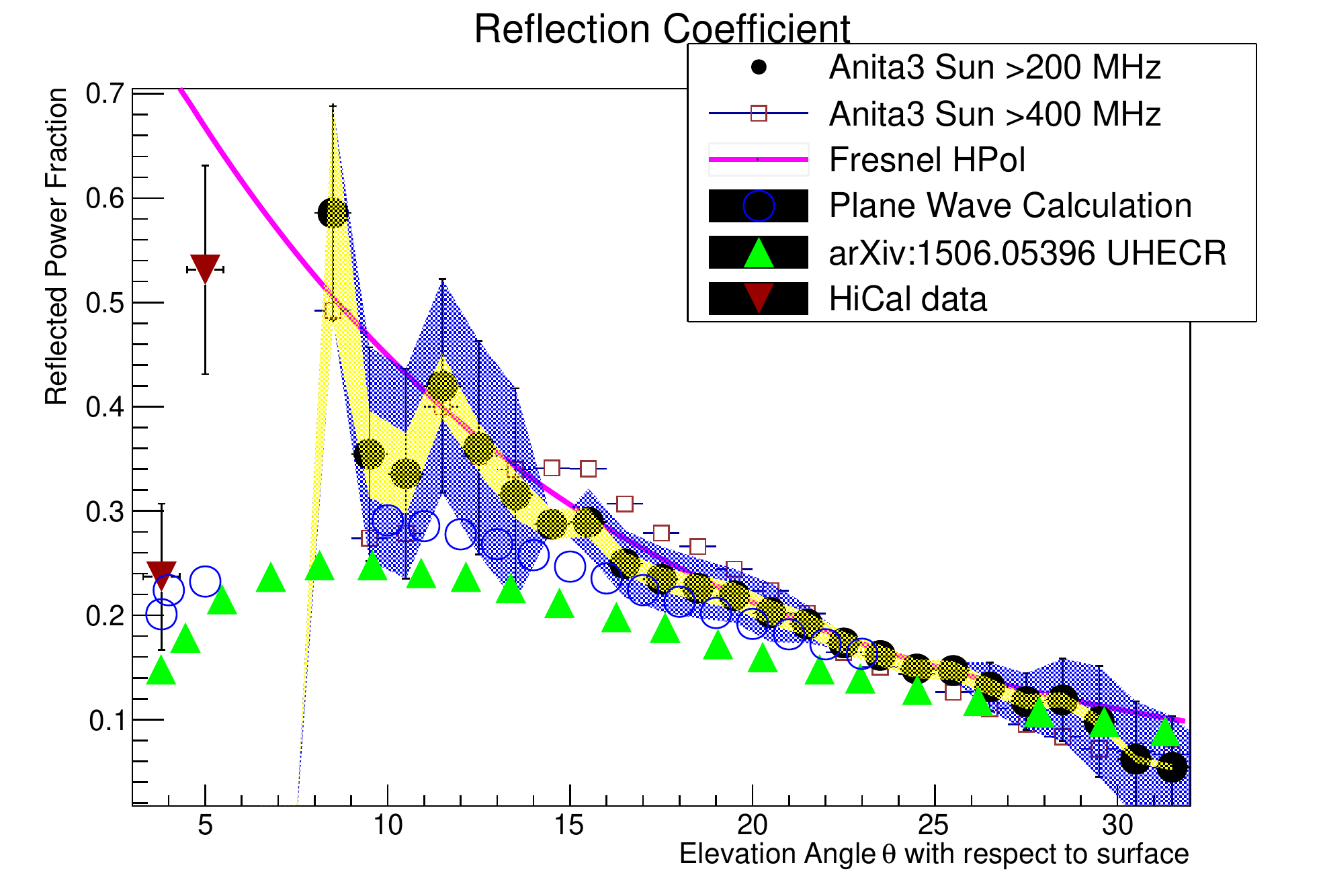}
\caption{\it Summary of ANITA-3 Antarctic radio frequency surface reflectivity measurements. Curve is Fresnel power reflection coefficient, assuming surface index-of-refraction of 1.35. Filled black circles are ANITA-3 Solar observations over full band 200-1000 MHz with blue bands and yellow bands indicating estimated uncertainties due to antenna beam width uncertainties and interferometric fringing effects, respectively; open brown squares show Solar reflectivity result, after high-pass filtering above 400 MHz. Inverted red-brown triangles show results obtained using HiCal-1b triggers observed by ANITA-3. Filled green triangles show currently applied corrections to UHECR energys; open blue circles show results of calculation described herein.}\label{fig:overlay.pdf} 
 \end{figure}

The value of the reflection coefficient can thus be inferred either directly from the scale factor needed
to `boost' the R-waveform to match the D-waveform, in voltage, or, alternately, by a direct measurement
of the received D vs. R power in an interferogram, which measures signal strength in units of $voltage^2$.
Compiling those two and combining with the Solar measurements, we present our measurements, along with
comparison to calculation (detailed below) and also the raw Fresnel coefficients, as shown in Figure \ref{fig:overlay.pdf}. Although the Solar measurements give good consistency with `raw' Fresnel, the HiCal measurements show a
clear deficiency of signal relative to the expectation from the Fresnel equations at oblique incidence angles $<5^\circ$.



\section{Estimates of Expected Reflectivity}
Given the indices of refraction of two dielectric media, the Fresnel coefficients for reflection and
transmission of power are derivable by imposing continuity at a smooth interface, taking into account the
difference in wavespeeds in the two media. These equations numerically prescribe the fraction of power
reflected by, and transmitted across that interface, assuming that all the scattering occurs at that 
interface. In practice, of course, there is some penetration across the dielectric boundary and into the dielectric, such that the actual reflected signal includes some sub-surface reflected signal.

\subsection{Reflection considerations}
In the most simple-minded approach, and neglecting roughness, we can approximate the Earth as a convex mirror, with focal length equal to half the Earth radius. At normal incidence, in the case where the object distance $d_s$ is quasi-infinite (the sun), the image distance is essentially the focal length, and the reflectance dictated by the Fresnel coefficients. If the object distance is much smaller than the focal 
length (the case for, e.g., either HiCal signals or radio-frequency signals from UHECR), now the image distance is 
approximately equal to the object distance.
If parallel rays traveling toward a convex mirror are not parallel to the main, there is still a 'virtual'
focal point $f$, although now $f\ne R_E/2$. In this case, the deviation from normal incidence is set by the
scale of the ratio of one Fresnel zone, projected at non-normal incidence, to the radius of curvature of the Earth.
At $\lambda\sim$1 m, and an observation distance $d_r$ of order 100 km from the reflection point, one Fresnel zone is of order
$\sqrt{\lambda d_sd_r/(d_s+d_r)}\sim$300 m, so the curvature correction is essentially negligible compared to the
radius of curvature.

A more sophisticated approach gives
the reduction in net power at the receiver as 
the reduction in flux density due to curvature, which can be
estimated by geometric considerations (this argument, of course, lacks a proper
accounting for phase variations which must be included). 
We can probe this possible effect experimentally by examining the
width of the interferometric peak for HiCal-1 data recorded at 3 degree elevation angle vs. 
4 degree incident elevation angle. However, within
our limited statistics, we find that these two are consistent with each other. 

Alternately, we can estimate the effect of reflection from a sphere of radius R, with source
at distance $d_s$ and elevation angle $\theta_s$ from the specular point 
and receiver at $d_r$ and $\theta_r$. The curvature of the reflecting surface results in
an `obscuration' factor describing the fraction of the total possible reflecting 
Antarctic surface which remains visible from either HiCal (in transmission) or ANITA
(in reception); we calculate this factor to be 61\% at $5^o$ 
incidence and 56\% at
$3.8^o$ incidence.
However, we stress that this is an extreme case,
corresponding to the entire visible surface contributing to the HiCal signal observed by ANITA.
We note that, although the signal is likely much greater than one Fresnel zone (in this case, of order
200 m), the limited time duration of the observed signals indicates a reflection area contributing to the
detected signal of order 50 km or so in radius only.












\section{Calculations}
We now present our calculations of the expected radio-frequency albedo,
using the Kirchoff approximation. A more rigorous treatment 
relies on the decomposition of a 
spherical wave into plane waves \cite{stratton1941electromagnetic}. 
For each plane wave the reflected 
and transmitted wave is given by the standard boundary conditions 
which lead to Fresnel coefficients. Integration over all the plane
waves gives the final electric field. 
This formalism turns out to be rather complicated for the case of 
a curved surface, such as the surface of Earth. Hence we postpone it 
to a future publication and confine ourselves here to the Kirchoff
scattering theory, which is also expected to be fairly reliable. 

\subsection{Scattering Theory: Calculation of the Kirchoff Integral}

We are interested in the received field strength $E_{rcv}$, which can be
expressed as the integral \citep{SWORD}  
\begin{equation}
E_{rcv} = {k\over 2\pi i} \int_{surface} E_{src}(\omega,\theta) {\cal F}(\theta)
F_{rough}(k,\rho,\theta){e^{ik(r+r')}\over r r'}\cos\theta dA
\label{eq:kirchoff}
\end{equation}
where $k$ is the wave number, ${\cal F}$ is the Fresnel coefficient and $F_{rough}$ is the correction
due to surface roughness.
 We integrate over the surface area and 
$\vec\rho$ is the position vector of the  integration point $Q$
relative to the specular point $O$, see
Fig. \ref{fig:kirchoff}. 
Here $\theta$ is the angle between the vector $\vec r\,'$ and the local 
normal to the surface at the point $Q$, i.e. the unit vector along 
$\vec R'$. 
We choose Cartesian coordinates $x,y,z$ on the tangent plane at $O$. 
In the figure $P'$ and $P$ represent the emitter and the receiver respectively.
The points $P$ and $P'$ both lie on the $y-z$ plane and the origin
$O$ is taken to be specular point. 
 The integral involves factors
such as $e^{ikr}/r$ and $e^{ikr'}/r'$, where $\vec r = \vec R_2-\vec \rho$
and $\vec r\,' = \vec R_1-\vec \rho$. The vector $\vec \rho= \vec R'-\vec R$.
Hence
\begin{equation}
\vec \rho = R_e\sin\theta'\cos\phi \hat i + R_e\sin\theta'\sin\phi \hat j 
+R_e(\cos\theta'-1)\hat k
\end{equation}  
where $R_e$ is the radius of the Earth and $\theta',\phi$ are the
standard spherical polar coordinates. 
It is convenient to define the coordinate $\rho_\perp=R_e\sin\theta'$. It is 
equal to projection of the vector $\vec \rho$ on the $x-y$ plane. 
Hence we obtain
\begin{equation}
\vec \rho = \rho_\perp\cos\phi \hat i + \rho_\perp\sin\phi \hat j 
+R_e(\cos\theta'-1)\hat k
\end{equation}  
with $\cos\theta' = \sqrt{1-\rho_\perp^2/R_e^2}$. 
We need to integrate over the surface of the sphere and hence the integration
measure is $R_e^2\sin\theta' d\theta' d\phi$. In terms of $\rho_\perp$ this becomes
$\rho_\perp d\rho_\perp d\phi/\sqrt{1-\rho_\perp^2/R_e^2}$. In terms of the Cartesian
coordinates, $x=\rho_\perp\cos\phi$, $y=\rho_\perp\sin\phi$, this simply becomes
$dxdy/\sqrt{1-\rho_\perp^2/R_e^2}$, with $\rho_\perp^2 = x^2+y^2$.  

The Fresnel coefficient is standard\citep{SWORD}. 
We have for the two polarization
states
\begin{eqnarray}
{\cal F}_\perp &=&  {
n_1\cos\theta - n_2\sqrt{1-\left({n_1\over n_2}\sin\theta\right)^2}\over 
n_1\cos\theta + n_2\sqrt{1-\left({n_1\over n_2}\sin\theta\right)^2}} 
\,, \nonumber\\ 
{\cal F}_\parallel &=&  {
n_2\cos\theta - n_1\sqrt{1-\left({n_1\over n_2}\sin\theta\right)^2}\over 
n_2\cos\theta + n_1\sqrt{1-\left({n_1\over n_2}\sin\theta\right)^2}} \,, 
\end{eqnarray}
which correspond to HPol and VPol respectively.
Here $n_1$ and $n_2$ are the indices of refraction of air and ice
respectively.

\subsection{Stationary Phase Approximation}
The phase in the integrand is a rapidly varying function of $x$ and $y$. 
Hence it is reasonable to use the stationary phase approximation. We expand
the phase around the stationary point, which is the same as the specular 
point and keep terms up to second order in the expansion. The coefficient
is assumed to be a slowly varying function of $x$ and $y$. In this case
it can be approximated by its value at the stationary point. We perform
the remaining integral keeping terms only up to second order in the phase. 
For example, the integral \cite{mandel1995optical} 
\begin{equation}
F(k) = \int dx dy f(x,y) e^{ikg(x,y)} \approx {2\pi i\over k\sqrt\Delta}
f(x_1,y_1) e^{ikg(x_1,y_1)}
\label{eq:stationary}
\end{equation}
where $(x_1,y_1)$ is the stationary point,  $\Delta = g_{xx}g_{yy}-g_{xy}^2$
and $g_{xx} = \partial^2g(x,y)/\partial x^2$, evaluated at the stationary
point, with analogous definitions for $g_{yy}$ and $g_{xy}$. 
Here we have assumed that $\Delta>0$ and $g_{xx}+g_{yy}>0$. 

The position vectors of the source and the receiver can be expressed as,
\begin{eqnarray}
\vec R_1 = y_1 \hat j + z_1\hat k\nonumber\\
\vec R_2 = y_2 \hat j + z_2\hat k
\end{eqnarray}
At the specular or stationary point
\begin{equation}
{y_1\over R_1} + {y_2\over R_2} = 0\,.
\end{equation}
We obtain
\begin{eqnarray}
r' &=& (R_1^2+\rho^2-2\vec R_1\cdot \vec \rho)^{1/2}\nonumber\\
r &=& (R_2^2+\rho^2-2\vec R_2\cdot \vec \rho)^{1/2} \,.
\end{eqnarray}
We need to expand $r+r'$ about the stationary point, keeping terms up to second
order in $x$ and $y$. We obtain
\begin{eqnarray}
r' &=& R_1\left[1-{y_1y\over R_1^2} + {1\over 2}{x^2+y^2\over R^2_1}
\left(1+{z_1\over R_e}\right)-{1\over 2}{y_1^2y^2\over R_1^4}\right]\,,\nonumber\\
r &=& R_2\left[1-{y_2y\over R_2^2} + {1\over 2}{x^2+y^2\over R^2_2}
\left(1+{z_2\over R_e}\right)-{1\over 2}{y_2^2y^2\over R_2^4}\right]\,.
\label{eq:rrp}
\end{eqnarray}
Using this we obtain
\begin{eqnarray}
g_{xx} &=& \left({1\over R_1} + {1\over R_2}\right) + {1\over R_e}
\left({z_1\over R_1} + {z_2\over R_2}\right)\,,\nonumber\\ 
g_{yy} &=& \left({1\over R_1} + {1\over R_2}\right)\cos^2\theta_0 + {1\over R_e}
\left({z_1\over R_1} + {z_2\over R_2}\right)\,,
\end{eqnarray}
and $g_{xy} = 0$. Hence $\Delta = g_{xx}g_{yy}$.  
We next evaluate the integral in Eq. \ref{eq:kirchoff} using 
Eq. \ref{eq:stationary}. The coefficient is evaluated at the stationary 
point. Hence we obtain 
\begin{equation}
E_{rcv} = {\cos\theta_0 \over \sqrt{\Delta}R_1R_2} 
E_{src}(\omega,\theta_0) {\cal F}(\theta_0)
F_{rough}(k,0,\theta_0)
\label{eq:kirchoff1}
\end{equation}
An overall factor $e^{ik(R_1+R_2)}$, which will not contribute to the power,
has been dropped. 
For an unpolarized beam the Fresnel coefficients lead to the factor
\begin{equation}
{\cal F}(\theta_0) = \sqrt{{1\over 2}\left({\cal F}^2_\perp+
{\cal F}^2_\parallel\right)}
\end{equation}
in the amplitude.

\begin{figure}[htpb]\begin{minipage}{0.4\textwidth}\centerline{\includegraphics[width=0.95\textwidth]{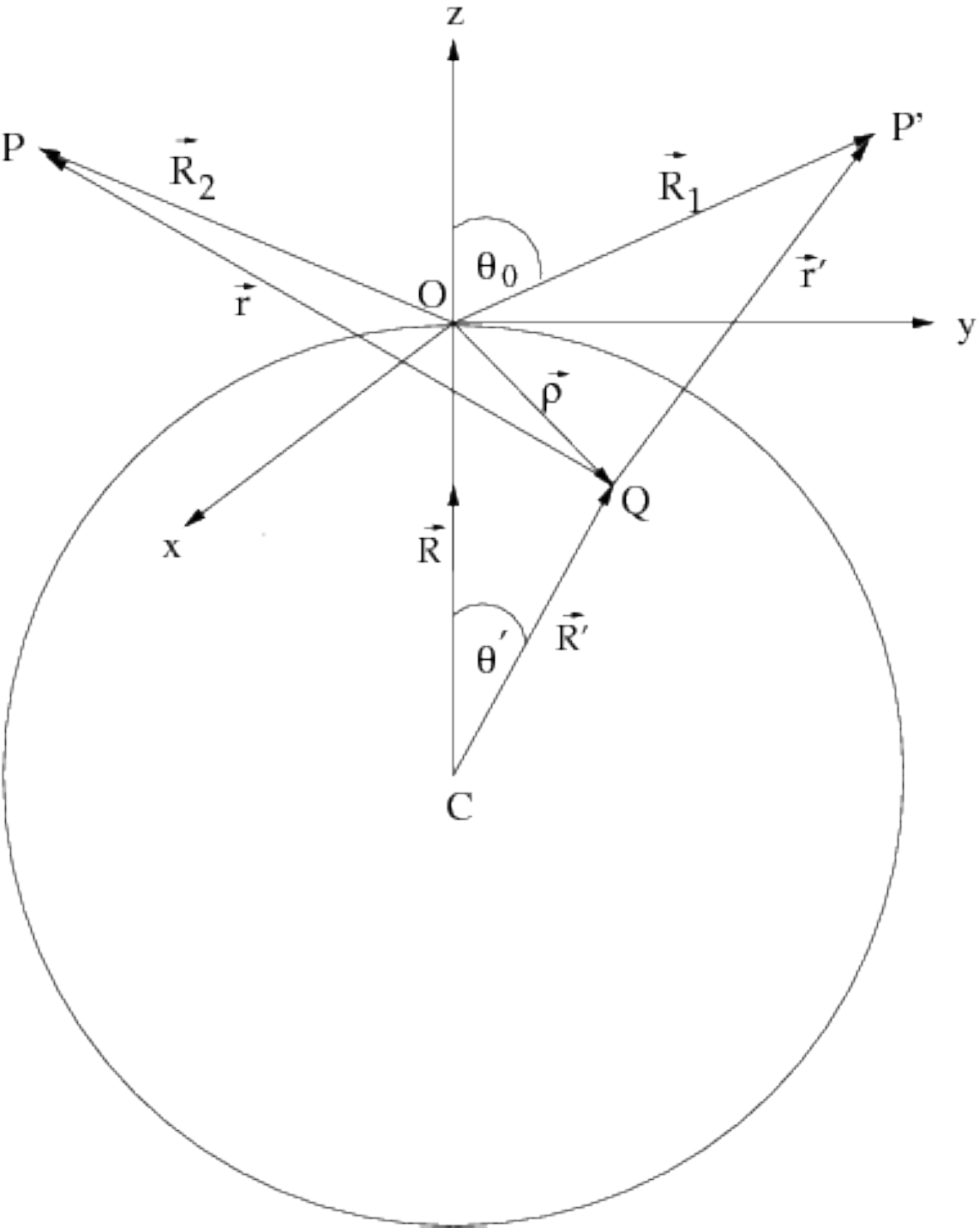}}\caption{\it The source is located at $P'$ and the receiver at $P$. The origin $O$ is chosen at the specular point such that angle of incidence is equal to that of reflection.}\label{fig:kirchoff}\end{minipage}
\begin{minipage}{0.48\textwidth}\centerline{\includegraphics[width=0.95\textwidth]{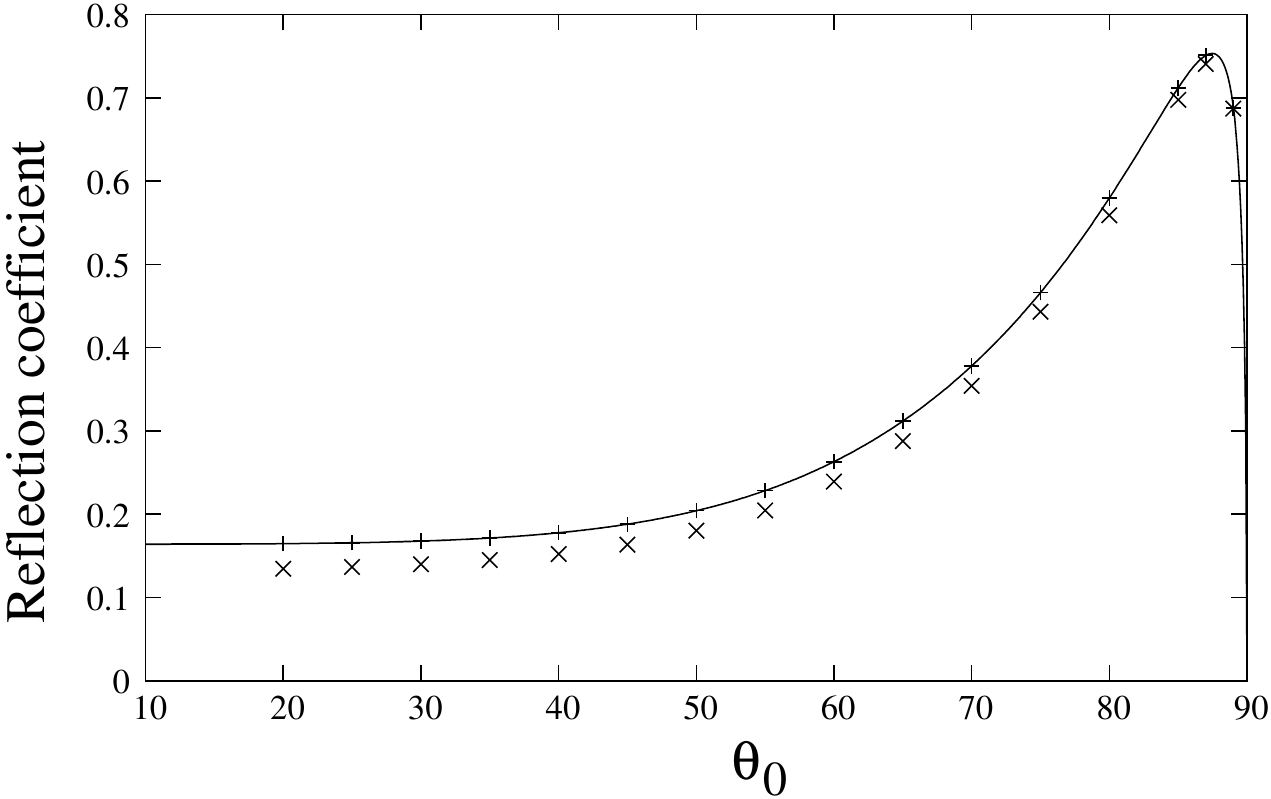}}\caption{\it The amplitude reflection coefficient as a function of the angle of incidence $\theta_0$. The solid curve represents the stationary phase approximation including the contribution due to Earth's curvature. The direct numerical result ($+$) is also shown for comparison. The crosses ($\times$) show the numerical result after including the roughness correction.}\label{fig:reflec}\end{minipage}\end{figure}

The overall factor $\cos\theta_0/(\sqrt{\Delta} R_1R_2)$ is found to
be equal to $\cos\theta_0/[(R_1+R_2)\delta_1\delta_2]$, where,
\begin{eqnarray}
\delta_1 &=& \left(1+2\cos\theta_0 {R_1R_2\over R_e(R_1+R_2)}\right)^{1/2}\,,\nonumber\\
\delta_2 &=& \left(\cos^2\theta_0+2\cos\theta_0 {R_1R_2\over R_e(R_1+R_2)}\right)^{1/2}\,.
\end{eqnarray}
Hence in the flat Earth limit, it is equal to $1/(R_1+R_2)$. 
We, therefore, find that the correction due to curvature is $\cos\theta_0/(\delta_1\delta_2)$. We plot the resulting amplitude reflectance 
coefficient as a function of the angle of incidence $\theta_0$
in Fig. \ref{fig:reflec}. Here we have set the index of refraction
of ice $n_2=1.4$ and have shown the result for an unpolarized beam.  
The altitudes of both the source and receiver has been set equal to 100 Km.

\subsection{Numerical Integration}
A direction numerical integration of the integral in Eq. \ref{eq:kirchoff}
over the $x,y$ variables is time consuming due to rapid fluctuations
of the integrand. Here we make a useful change of variables motivated by
the stationary phase approximation which helps in speeding up the
integral evaluation. We notice that the dominant contribution arises from the 
terms quadratic in $x$ and $y$ in the exponent $r+r'$. The relevant terms
can be obtained from Eq. \ref{eq:rrp}. We can express these terms as
$${1\over 2}g_{xx} x^2 + {1\over 2}g_{yy}y^2\,.$$ 
Notice that the terms linear in $x,y$ drop out in the stationary phase
approximation. We now use the scaled 
variables
\begin{equation}
\tilde x = \sqrt{g_{xx}}\, x \,,\ \ \ \ \ \ \ 
\tilde y = \sqrt{g_{yy}}\, y\,.
\end{equation}
In terms of these the quadratic part is proportional to $\tilde x^2+\tilde y^2$.
We next use the plane polar variables $\tilde\rho$, $\tilde\phi$,
such that, 
\begin{eqnarray}
\tilde x &=& \tilde \rho \cos\tilde \phi\nonumber\\
\tilde y &=& \tilde \rho \sin\tilde \phi
\end{eqnarray}
It is clear that in terms of these variables the dominant part of the
integral becomes essentially one dimensional, i.e. the integrand
has very weak dependence on $\tilde \phi$. Hence we can now make a 
numerical calculation by choosing a dense grid in the variable $\tilde\rho$
and a relatively sparse grid in $\tilde \phi$. We find that this leads
to a significant gain in convergence speed. The result of numerical
integration, assuming a monochromatic source of wavelength equal to 1 m, 
is also shown in Fig. \ref{fig:kirchoff}. We find that it
shows excellent agreement with the analytic approximation. 
We point out that in the stationary phase approximation the result is
found to be independent of frequency. Hence the full numerical estimation
is also not expected to show a significant dependence on frequency.

\subsection{Surface Roughness}
We next include the surface roughness contribution using the method 
described in \cite{SWORD}. The roughness factor $ F_{rough}(k,\rho,\theta)$ 
is modelled as,
\begin{equation}
 F_{rough}(k,\rho,\theta) = \exp\left[{-2k^2\sigma_h(\rho_\perp)^2\cos^2\theta_0}\right]
\end{equation}
where $\rho_\perp^2 = x^2+y^2$ and  
\begin{equation}
\sigma_h(L) = \sigma_h(L_0)\left({L\over L_0}\right)^H\,.
\end{equation}
We choose the parameters $L_0=120$ m, $\sigma_h(120 m) = 0.04$ m 
and $H=0.65$. The result for the case of frequency $\nu = 300$ MHz,
  obtained by direct numerical integration, is shown
in Fig. \ref{fig:reflec}. 
In this case the stationary phase approximation does not appear to  be
reliable since the exponential factor appears to show significant 
dependence on the integration variables.

\begin{wrapfigure}{R}{0.45\textwidth}
\includegraphics[width=0.45\textwidth]{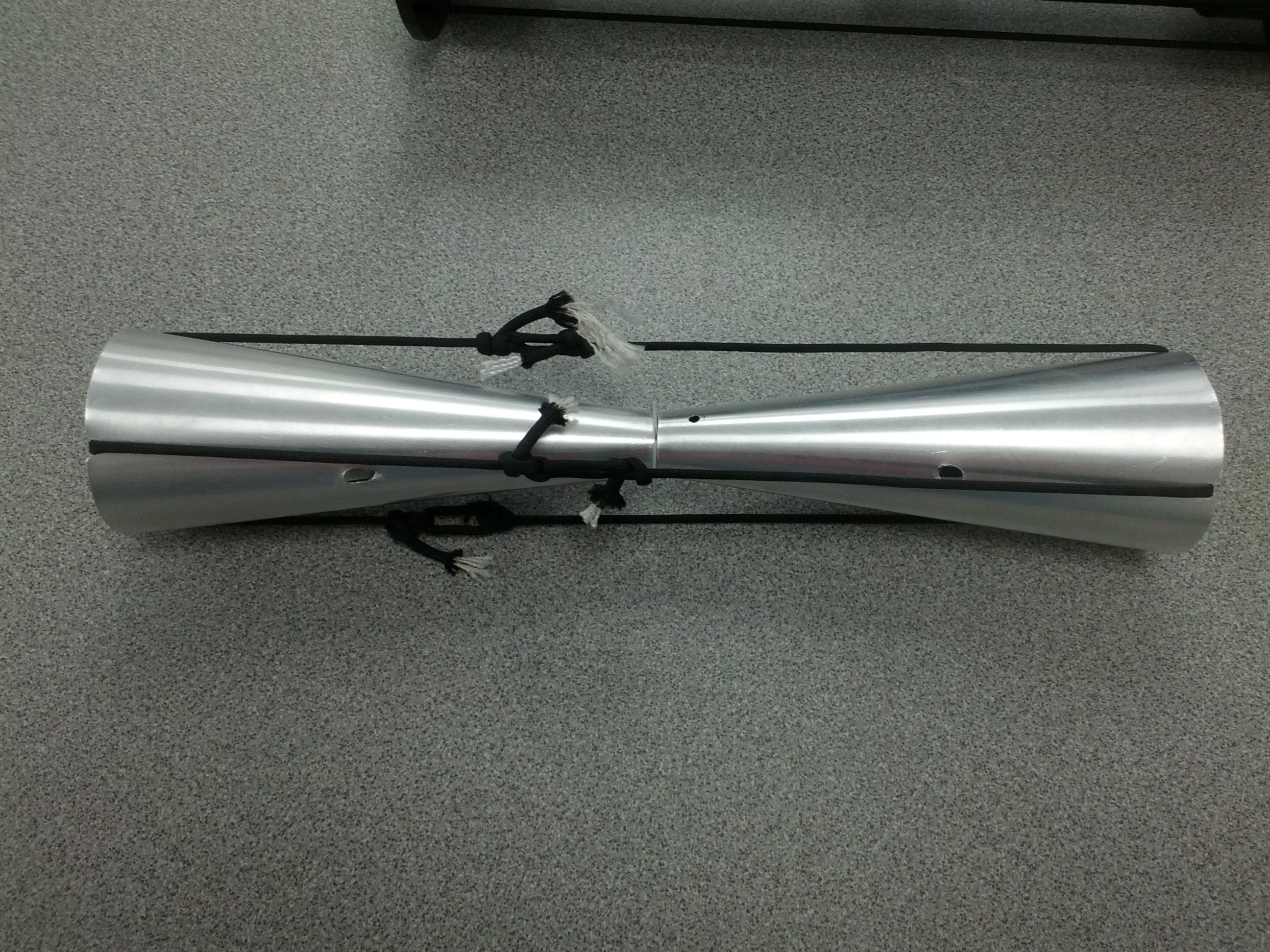}
\caption{\it Photograph of HiCal-2 transmitter antenna.} \label{fig:hical2-Tx.jpg} 
\end{wrapfigure}

\section{The HiCal-2 mission}
Two HiCal payloads were successfully launched in December, 2016 (``HiCal-2a'' and ``HiCal-2b''). These both featured two piezos on each transmitter antenna (for redundancy), with the discharge from each piezo directed across the antenna feedpoint. 
Other improvements for HiCal-2 and future versions include:

\begin{itemize}
\item Local readout of each emitted transmitter signal amplitude using a simple 3-bit comparator. \message{HiCom}
A simple scheme consisting of several tiers of comparators to provide an estimate of the pulse-to-pulse piezo signal output strength is adequate here. Relative to HiCal-1, which showed considerable pulse-to-pulse signal variation, HiCal-2 should be considerably more stable. Note that, with improved resolution, one can, in principle, perform a direct 'mono-static' measurement of the surface reflectivity with a surface receiver directly mounted to HiCal (``Ultra-Cal'').

\item Local readout of GPS-times of transmitter signals - this has already been achieved for HiCal-1, although the triggering of the GPS readout was somewhat brittle and poor isolation of the trigger board often resulted in saturation of the trigger board electronics. A simple scheme, whereby the piezo trigger is tapped directly into a GPS unit is our primary candidate for HiCal-2 implementation.
For the HiCal-2 flight, our custom board was updated to provide, in additional to Azimuthal information, a GPS time stamp with resolution of 30 $\mu$s; this resolution is primarily determined by the clock speed used to latch the GPS time. For this upgraded board (``ATSA''), the photodiodes have been replaced by silicon photomultipliers designed and assembled at the Moscow Engineering Physics Institute (MEPhI), achieving similar sub-degree angular resolution.


\end{itemize}

\begin{figure}[htpb]
\begin{minipage}{7in}
\centerline{\includegraphics[width=1.\textwidth]{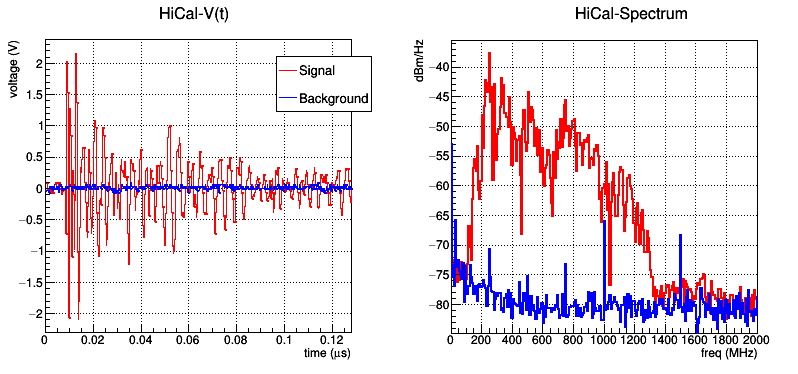}} 
\caption{\it Signal output in the time domain and also frequency domain power spectrum, with background overlaid. Signal is measured at a distance of 30 meters from an ANITA-2 horn antenna, with no additional front-end amplification.} \label{fig:hical2-signalFFT.png} \end{minipage} \end{figure}

The HiCal-2 antenna is shown in Figure \ref{fig:hical2-Tx.jpg}; 
in this case, the piezo signal is directed to, and
discharged across a small gap at the feedpoint of the `ice-cream cone' antenna. 
Our studies thus far indicate that the maximum signal strength is obtained by minimizing the gap across the two antenna halves. Our default configuration uses a 300 micron spark gap;
Figure \ref{fig:hical2-signalFFT.png} demonstrates an adequately broadband signal.

\section{Summary}
At frequencies above 2 GHz, satellite data show generally consistent reflected power maps, as well as
correlations of reflectivity with wind speed over the bulk of the microwave frequency regime. 
In the ANITA passband (200-1000 MHz), we have performed measurements of the Solar surface-reflection,
covering incident angles above 10 degrees. A formalism has been developed which, within errors,
reproduces the Solar data.
The HiCal experiment was designed to measure the Antarctic surface reflectivity at oblique angles.
At frequencies below
1 GHz, and at very glancing incidence angles, our HiCal-1 measurements indicate important surface-decoherence effects. The December, 2016 HiCal-2 mission increased, by over an order of magnitude, the statistics obtained from HiCal-1, and, provided reflectivity measurements over a much broader range of incidence angles.

\section{Acknowledgments}
We thank NASA for their generous support of ANITA and HiCal, 
the  Columbia  Scientific  Balloon  Facility  for  their  excellent
field support, and the National Science Foundation for their
Antarctic operations support.  This work was also supported
by the US Dept. of Energy, High Energy Physics Division, 
as well as NRNU, MEPhI and 
 the Megagrant 2013 program of Russia, via agreement 14.А12.31.0006 from 24.06.2013.
\bibliography{/home/dbesson/updoc/Zref.bib}
\bibliographystyle{unsrt}

\end{document}